**Enrico Garaldi** [1]

**1** Max-Planck-Institüt für Astrophysik, Karl-Schwarzschild-Str. 1, 85741 Garching, Germany

## Summary

The number of available constraints on the Universe during and before cosmic reionization is rapidly growing. These are often scattered across inhomogeneous formats, unit systems and sampling strategies. In this paper, I introduce CoReCon, a Python package designed to provide a growing set of constraints on key physical quantities related to the Epoch of Reionization and a platform for the high-redshift research community to collect and store, in an open way, current and forthcoming observational constraints.

## Statement of Need

The Epoch of Reionization (EoR) is the last global phase transition in the history of the Universe, and it represents the current frontier in the study of galaxy formation, as it radically altered the environment in which cosmic structures formed and evolved thereafter. It transformed the intergalactic medium (IGM) that fills the space between galaxies from a cold and neutral gas to a hot and ionized one (for a review see Wise, 2019). Despite its importance, little is known about this period of time, mostly as a consequence of the intrinsic difficulties in observing such a distant epoch. This is now rapidly changing thanks to the *James Webb Space Telescope*, which is providing exquisite observations of the high-redshift Universe (see e.g. Chen et al., 2022; Curti et al., 2022; Donnan et al., 2022; Endsley et al., 2022; Finkelstein et al., 2022; Furtak et al., 2022; Harikane et al., 2022; Laporte et al., 2022; Leethochawalit et al., 2022; Naidu et al., 2022; Roberts-Borsani et al., 2022).

To gain some insight into the EoR, a plethora of different methods have been devised to extract information from the limited observations available. However, these data are typically scattered in many different publications, using inhomogeneous unit systems, and sampling strategies (e.g. volume- or mass-averaged quantities in the intergalactic medium). Hence, employing these data in a scientifically-sound way often requires (i) retrieving the aforementioned information from different publications, and (ii) homogenizing them. The situation is made worse by the fact that derived constraints are often not explicitly reported (although this is slowly changing), forcing their retrieval from published *plots* when the authors are unavailable (e.g. because they have moved on to a different career) or unwilling to share the data, a tedious and error-prone task and a substandard scientific practice.

I tackle these issues through a systematic collection of published constraints on the physical properties of the Universe during the EoR. This collection is wrapped in a Python module named CoReCon (an acronym for Collection of Reionization Constraint) that I present in this paper. I start by introducing the goals and design choices of CoReCon, then move on to a description of its features, followed by a review of the available constraints at the time of publication. I close this paper by discussing desirable future developments and with a pledge to the community.

## Overview

The goal of `CoReCon` is twofold. First, it comprises a growing set of constraints on key physical quantities related to the EoR, homogenized in their format and units, lifting the busy researchers from the burden of searching, retrieving and formatting data. Second, and foremost, `CoReCon` provides a platform for the high-redshift research community to collect and store, in an open way, such observational constraints. I do so by providing a Python infrastructure, which is able to load formatted data files and provides simple utility functions to deal with such data. The data files loaded by `CoReCon` are purposely simple in their form and as complete as possible in their content, in order to collect all the relevant information in one place. Notably, they are required to contain a URL to the original publication and a short description of the methods used to retrieve the constraints from observed data. They are also allowed to contain any additional, unplanned-for data field, in order to reach the highest degree of flexibility and to allow the storage of all relevant information. Ideally, once a new constraint is published, the author of the publication will update `CoReCon` with the relevant data. If this procedure becomes customary, `CoReCon` will serve as an up-to-date repository of easy-to-retrieve constraints.

To our knowledge, this is the first module of its kind – at least in the EoR community. With a similar spirit, there exist a collection of all the known quasars above a redshift of 5.7 (Bosman, 2020) and a compilation of galaxy data for the specific purpose of validating the `VELOCIraptor` halo finder (The SWIFTSIM team, 2022). Additionally, an effort toward openness of research in the EoR field recently materialized into an open analysis pipeline for the reduction of spectra taken in most of the major telescopes in the world (Prochaska et al., 2020).

In its development version, `CoReCon` has been used for the scientific analysis of the THESAN simulations (Garaldi et al., 2022a; Kannan et al., 2022; Smith et al., 2022) and is being used in upcoming scientific projects.

## Features

`CoReCon` is written as a Python module in order to provide portability, ease of installation and use, and to reach the large community of researchers using Python. Additionally, I put effort into building a template for entering new data into the module, which strives to be simultaneously easy to fill and complete in its content.

The `CoReCon` module is able to read two different data layouts and internally transforms them into the frontend data format exposed to the user. The module also includes simple utility functions that can transform the available data in commonly-used ways. For instance, selecting only the constraint on a specific physical quantity, in a user-defined redshift range, or transforming their layout to be ready-to-plot using the `matplotlib` Python module.

`CoReCon` has been developed with openness in mind. For this reason, new constraints can be easily added by simply filling a form provided, and copying it into the directory tree of the module. Data entries are required to contain the reference and a link to the original publication, in order to ensure the original publication is acknowledged, a *quality* flag which specifies if the data were explicitly available in the publication or has been retrieved in some indirect way (e.g. from a published plot, hence potentially introducing errors[1]), and a short description of the constraints themselves and of the method employed to measure/compute them.

The `CoReCon` module can be easily installed via `pip` and is fully documented online at https://corecon.readthedocs.io/en/latest/. CoReCon autonomously fetches updates to the constraints at startup (but limited to once every 24 hours or when manually triggered to do so by the user), in order to remove the requirement to manually update the entire package to obtain new constraints.

---

[1] While ideally I would like to simply ignore the data that are not available through the published paper or the authors themselves, this would limit significantly the number of constraints available. In addition, in many cases, the retrieved data are quite faithful to the original values. I have tested this by retrieving data from plots in publications that also reported the numerical values, and compared the two. Finally, I notice that I provide the option of filtering the data based on their retrieval method, in order to leave the users the freedom to choose which constraints to rely on.

Finally, the CoReCon repository features continuous integration through GitHub Actions, ensuring each new commit is tested and satisfies a minimal functionality level.

### Technical implementation

The main structures in CoReCon are the Field and DataEntry classes, respectively representing a collection of constraints on a single physical quantity and the constraints from an individual source (as a scientific publication). These are supplemented by a custom data format for the storage of the data.

The Field class is inherited from python's native dictionary class, and enriches the latter with additional variables describing the physical quantity represented as well as its units, commonly-adopted scientific symbols and important remarks. This approach was chosen in such a way to isolate the class members corresponding to individual constraint entries, which are represented by the dictionary keys, and members describing the physical quantity as a whole, which are non-keys class members. Additionally, this allowed us to include utility functions (as e.g. filter functions to select constraints based on custom criteria) in the class itself, allowing for an easy concatenation of them.

Individual constraints are implemented through the custom DataEntry class and the corresponding data format for their storage. The latter is thoroughly described online at https://corecon.readthedocs.io/en/latest/, while the former takes care of loading the data, checking their format and expanding (when possible) fields in the native format of CoReCon.

### Available constraints

At the time of writing, CoReCon contains data for the following physical quantities:
- *ionized fraction*. The ionized fraction of hydrogen in the Universe. Notice that this contains both volume-averaged (the majority) and mass-averaged values. The type of average is detailed in the description of each dataset.
- *IGM temperature at mean density*. This value is typically model-dependent, as its derivation involves calibration against simulations.
- *effective optical depth of the HI and HeII Lyman-$\alpha$ forest*.
- *flux power spectrum of the Lyman-$\alpha$ forest*.
- *cosmic microwave background optical depth*.
- *UV luminosity function*. I provide the logarithm of this value and the associated errors.
- *quasar luminosity function*. - *column density ratio of HeII to HI*.
- *mean free path of ionizing photons*.
- *star-formation-rate density evolution*.
- *average transmitted flux quasar spectra as a function of distance from nearby galaxies*. This provides information on the sources (Kakiichi et al., 2018; R. A. Meyer et al., 2019; Romain A. Meyer et al., 2020) and timing (Garaldi et al., 2022b) of reionization.
- *mass-metallicity relation of galaxies*, both for stellar and gas-phase metallicities.
- *galaxy main sequence* (i.e. star formation rate as a function of stellar mass).
- *UV slope*, defined as the slope of the flux density in wavelength blueward of the Lyman-$\alpha$ line.

I provide a small description of each field within CoReCon itself, as a string attached to each set of constraints.

The full list of available constraints is constantly updated. Therefore I refer the reader to the relative documentation page.

# Future work

By its nature, CoReCon is an ever-evolving package. Not only new constraints will constantly be published, but new instruments and techniques will enable the observations of new physical quantities. I will update CoReCon consequently to allow their inclusion. An example of possible improvement is the implementation of a new data structure representing an individual astronomical object, with one or multiple sources for its (inferred) physical properties. Relations between quantities could therefore be dynamically generated from (or complemented by) the collection of such objects. This provides a solution to the current issue that the same object may have multiple entries, one for each relation in which it appears (e.g. UV luminosity function and galaxy main sequence).

Another foreseen improvement is the integration of CoReCon with the pandas module (team, 2020), in order to return a pandas DataFrame when fetching a constraint. This will open up the possibility to employ the wide array of features available through pandas within CoReCon.

Finally, I plan to include informations about the cosmology and initial mass function assumed in deriving constraints, alongside functionalities to convert the data to a target cosmology or initial mass function.

# Pledge to the EoR community

CoReCon is an open and collaborative project of its own nature. I strongly believe it can be a useful tool for the EoR community, but it can only thrive and be so through a collaborative effort. I ask everyone that finds this module useful for their research to contribute and enrich the constraints collection, providing new entries.

# Acknowledgements

I acknowledge in advance the community that – I am sure – will help make CoReCon a complete and valuable module. I am grateful to Benedetta Ciardi, Martin Glatzle, Aniket Bhagwat and Adam Schaefer for useful comments, discussions and beta-testing. I am thankful to the community for developing and maintaining the numpy (Harris et al., 2020) software package, upon which CoReCon is built, and matplotlib (Hunter, 2007), employed for the data visualization in the CoReCon documentation.

# References

Bosman, S. E. I. (2020). *All z>5.7 quasars currently known* (Version 1.5) [Data set]. Zenodo. https://doi.org/10.5281/zenodo.3909340

Chen, Z., Stark, D. P., Endsley, R., Topping, M., Whitler, L., & Charlot, S. (2022). JWST/NIRCam Observations of Stars and HII Regions in $z \simeq 6 - 8$ Galaxies: Properties of Star Forming Complexes on 150 pc Scales. *arXiv e-Prints*, arXiv:2207.12657. https://arxiv.org/abs/2207.12657

Curti, M., D'Eugenio, F., Carniani, S., Maiolino, R., Sandles, L., Witstok, J., Baker, W. M., Bennett, J. S., Piotrowska, J. M., Tacchella, S., Charlot, S., Nakajima, K., Maheson, G., Mannucci, F., Arribas, S., Belfiore, F., Bonaventura, N. R., Bunker, A. J., Chevallard, J., … Wallace, I. E. B. (2022). The chemical enrichment in the early Universe as probed by JWST via direct metallicity measurements at z~8. *arXiv e-Prints*, arXiv:2207.12375. https://arxiv.org/abs/2207.12375

Donnan, C. T., McLeod, D. J., Dunlop, J. S., McLure, R. J., Carnall, A. C., Begley, R., Cullen, F., Hamadouche, M. L., Bowler, R. A. A., McCracken, H. J., Milvang-Jensen, B.,

Moneti, A., & Targett, T. (2022). The evolution of the galaxy UV luminosity function at redshifts z ~8-15 from deep JWST and ground-based near-infrared imaging. *arXiv e-Prints*, arXiv:2207.12356. https://arxiv.org/abs/2207.12356

Endsley, R., Stark, D. P., Whitler, L., Topping, M. W., Chen, Z., Plat, A., Chisholm, J., & Charlot, S. (2022). A JWST/NIRCam Study of Key Contributors to Reionization: The Star-forming and Ionizing Properties of UV-faint $z \sim 7-8$ Galaxies. *arXiv e-Prints*, arXiv:2208.14999. https://arxiv.org/abs/2208.14999

Finkelstein, S. L., Bagley, M. B., Arrabal Haro, P., Dickinson, M., Ferguson, H. C., Kartaltepe, J. S., Papovich, C., Burgarella, D., Kocevski, D. D., Huertas-Company, M., Iyer, K. G., Larson, R. L., Pérez-González, P. G., Rose, C., Tacchella, S., Wilkins, S. M., Chworowsky, K., Medrano, A., Morales, A. M., … Zavala, J. A. (2022). A Long Time Ago in a Galaxy Far, Far Away: A Candidate z ~14 Galaxy in Early JWST CEERS Imaging. *arXiv e-Prints*, arXiv:2207.12474. https://arxiv.org/abs/2207.12474

Furtak, L. J., Shuntov, M., Atek, H., Zitrin, A., Richard, J., Lehnert, M. D., & Chevallard, J. (2022). Constraining the physical properties of the first lensed $z \sim 10-16$ galaxy candidates with JWST. *arXiv e-Prints*, arXiv:2208.05473. https://arxiv.org/abs/2208.05473

Garaldi, E., Kannan, R., Smith, A., Springel, V., Pakmor, R., Vogelsberger, M., & Hernquist, L. (2022b). The THESAN project: properties of the intergalactic medium and its connection to reionization-era galaxies. *Monthly Notices of the Royal Astronomical Society*, *512*(4), 4909–4933. https://doi.org/10.1093/mnras/stac257

Garaldi, E., Kannan, R., Smith, A., Springel, V., Pakmor, R., Vogelsberger, M., & Hernquist, L. (2022a). The THESAN project: properties of the intergalactic medium and its connection to reionization-era galaxies. *Monthly Notices of the Royal Astronomical Society*, *512*(4), 4909–4933. https://doi.org/10.1093/mnras/stac257

Harikane, Y., Inoue, A. K., Mawatari, K., Hashimoto, T., Yamanaka, S., Fudamoto, Y., Matsuo, H., Tamura, Y., Dayal, P., Yung, L. Y. A., Hutter, A., Pacucci, F., Sugahara, Y., & Koekemoer, A. M. (2022). A Search for H-Dropout Lyman Break Galaxies at z 12-16. *The Astrophysical Journal*, *929*(1), 1. https://doi.org/10.3847/1538-4357/ac53a9

Harris, C. R., Millman, K. J., Walt, S. J. van der, Gommers, R., Virtanen, P., Cournapeau, D., Wieser, E., Taylor, J., Berg, S., Smith, N. J., Kern, R., Picus, M., Hoyer, S., Kerkwijk, M. H. van, Brett, M., Haldane, A., R'ıo, J. F. del, Wiebe, M., Peterson, P., … Oliphant, T. E. (2020). Array programming with NumPy. *Nature*, *585*(7825), 357–362. https://doi.org/10.1038/s41586-020-2649-2

Hunter, J. D. (2007). Matplotlib: A 2D graphics environment. *Computing In Science & Engineering*, *9*(3), 90–95. https://doi.org/10.1109/MCSE.2007.55

Kakiichi, K., Ellis, R. S., Laporte, N., Zitrin, A., Eilers, A.-C., Ryan-Weber, E., Meyer, R. A., Robertson, B., Stark, D. P., & Bosman, S. E. I. (2018). The role of galaxies and AGN in reionizing the IGM - I. Keck spectroscopy of $5 < z < 7$ galaxies in the QSO field J1148+5251. *Monthly Notices of the Royal Astronomical Society*, *479*, 43–63. https://doi.org/10.1093/mnras/sty1318

Kannan, R., Garaldi, E., Smith, A., Pakmor, R., Springel, V., Vogelsberger, M., & Hernquist, L. (2022). Introducing the THESAN project: radiation-magnetohydrodynamic simulations of the epoch of reionization. *Monthly Notices of the Royal Astronomical Society*, *511*(3), 4005–4030. https://doi.org/10.1093/mnras/stab3710

Laporte, N., Zitrin, A., Dole, H., Roberts-Borsani, G., Furtak, L. J., & Witten, C. (2022). A lensed protocluster candidate at $z = 7.66$ identified in JWST observations of the galaxy cluster SMACS0723-7327. *arXiv e-Prints*, arXiv:2208.04930. https://arxiv.org/abs/2208.04930

Leethochawalit, N., Trenti, M., Santini, P., Yang, L., Merlin, E., Castellano, M., Fontana, A., Treu, T., Mason, C., Glazebrook, K., Jones, T., Vulcani, B., Nanayakkara, T., Marchesini, D., Mascia, S., Morishita, T., Roberts-Borsani, G., Bonchi, A., Paris, D., … Scarlata, C. (2022). Early results from GLASS-JWST. X: Rest-frame UV-optical properties of galaxies at 7 < z < 9. *arXiv e-Prints*, arXiv:2207.11135. https://arxiv.org/abs/2207.11135

Meyer, R. A., Bosman, S. E. I., Kakiichi, K., & Ellis, R. S. (2019). The role of galaxies and AGNs in reionizing the IGM - II. Metal-tracing the faint sources of reionization at 5 < z < 6. *Monthly Notices of the Royal Astronomical Society*, *483*, 19–37. https://doi.org/10.1093/mnras/sty2954

Meyer, Romain A., Kakiichi, K., Bosman, S. E. I., Ellis, R. S., Laporte, N., Robertson, B. E., Ryan-Weber, E. V., Mawatari, K., & Zitrin, A. (2020). The role of galaxies and AGN in reionizing the IGM - III. IGM-galaxy cross-correlations at z 6 from eight quasar fields with DEIMOS and MUSE. *Monthly Notices of the Royal Astronomical Society*, *494*(2), 1560–1578. https://doi.org/10.1093/mnras/staa746

Naidu, R. P., Oesch, P. A., van Dokkum, P., Nelson, E. J., Suess, K. A., Whitaker, K. E., Allen, N., Bezanson, R., Bouwens, R., Brammer, G., Conroy, C., Illingworth, G., Labbe, I., Leja, J., Leonova, E., Matthee, J., Price, S. H., Setton, D. J., Strait, V., … Weibel, A. (2022). Two Remarkably Luminous Galaxy Candidates at $z \approx 11-13$ Revealed by JWST. *arXiv e-Prints*, arXiv:2207.09434. https://arxiv.org/abs/2207.09434

Prochaska, J. X., Hennawi, J. F., Westfall, K. B., Cooke, R. J., Wang, F., Hsyu, T., & Farina, E. P. (2020). PypeIt: The Python Spectroscopic Data Reduction Pipeline. *arXiv e-Prints*, arXiv:2005.06505. https://doi.org/10.21105/joss.02308

Roberts-Borsani, G., Morishita, T., Treu, T., Leethochawalit, N., & Trenti, M. (2022). The Physical Properties of Luminous z ≳ 8 Galaxies and Implications for the Cosmic Star Formation Rate Density from 0.35 deg$^2$ of (Pure-)Parallel HST Observations. *The Astrophysical Journal*, *927*(2), 236. https://doi.org/10.3847/1538-4357/ac4803

Smith, A., Kannan, R., Garaldi, E., Vogelsberger, M., Pakmor, R., Springel, V., & Hernquist, L. (2022). The THESAN project: Lyman-$\alpha$ emission and transmission during the Epoch of Reionization. *Monthly Notices of the Royal Astronomical Society*, *512*(3), 3243–3265. https://doi.org/10.1093/mnras/stac713

team, T. pandas development. (2020). *Pandas-dev/pandas: pandas* (latest). Zenodo. https://doi.org/10.5281/zenodo.3509134

The SWIFTSIM team. (2022). *VELOCIraptor comparison data*. https://github.com/swiftsim/velociraptor-comparison-data.

Wise, J. H. (2019). An Introductory Review on Cosmic Reionization. *arXiv e-Prints*, arXiv:1907.06653. https://arxiv.org/abs/1907.06653